\documentclass[aps,pre,twocolumn]{revtex4}
\usepackage{graphics}% Include figure files
\usepackage{graphicx}% Include figure files

\begin{document}

%\preprint{}

%\title{Subdiffusive scaling laws for turbulence spreading by wave-wave interactions}
\title{Turbulence spreading by the resonant wave-wave interactions: A fractional kinetics approach}


\author{Alexander~V.~Milovanov${}^{1,2}$ and Jens~Juul~Rasmussen${}^{3}$}
%\email[]{Alexander.Milovanov@phys.uit.no}
%\homepage[]{www.phys.uit.no}
%\thanks{}
%\altaffiliation{Also at: Department of Space Plasma Physics, Space Research Institute, Russian Academy of Sciences, Profsoyuznaya 84/32, 117997 Moscow, Russia}

\affiliation{${}^1$ENEA National Laboratory, Centro~Ricerche~Frascati, I-00044 Frascati, Rome, Italy}
\affiliation{${}^2$Max-Planck-Institut f\"ur Physik komplexer Systeme, D-01187 Dresden, Germany}
\affiliation{${}^3$Physics Department, Technical University of Denmark, DK-2800 Kgs.~Lyngby, Denmark}

%\email[]{jens.juul.rasmussen@risoe.dk}
%\thanks{}


%\date{\today}

\begin{abstract} This paper is concerned with the processes of spatial propagation and penetration of turbulence from the regions where it is locally excited into initially laminar regions. The phenomenon has come to be known as ``turbulence spreading" and witnessed a renewed attention in the literature recently. Here, we propose a comprehensive theory of turbulence spreading based on fractional kinetics. We argue that the use of fractional-derivative equations permits a general approach focussing on fundamentals of the spreading process regardless of a specific turbulence model and/or specific instability type. The starting point is the Hamiltonian of the resonant wave-wave interactions, from which a family of scaling laws for the asymptotic spreading is derived. Both three- and four-wave interactions are considered. The results span from a subdiffusive spreading in the parameter range of weak chaos to avalanche propagation in regimes with population inversion. Attention is paid to how non-ergodicity introduces weak mixing, memory and intermittency into spreading dynamics, and how the properties of non-Markovianity and nonlocality emerge from the presence of islands of regular dynamics in phase space. Also we resolve an existing question concerning turbulence spillover into gap regions, where the instability growth is locally suppressed, and show that the spillover occurs through exponential (Anderson like) localization in case of four-wave interactions and through an algebraic (weak) localization in case of triad interactions. In the latter case an inverse-cubic behavior of the spillover function is found. Wherever relevant, we contrast our findings against the available observational and numerical evidence, and we also commit ourselves to establish connections with the models of turbulence spreading proposed previously.             
\end{abstract}

%\pacs{05.45.Mt, 72.15.Rn, 42.25.Dd, 05.45.-a}

\maketitle

\section{Introduction} 

Turbulence spreading$-$also known as turbulence penetration or turbulence overshoot$-$is the process of propagation and expansion of an initially localized puff of turbulence into surrounding areas. The phenomenon characterizes both fluid and plasma turbulence and has been examined for applications in solar and astrophysics \cite{Tao,Brummel}, geophysics \cite{Large,Plume} and magnetic confinement fusion \cite{Garbet,Itoh,Naulin}. The interest one pays to turbulence spreading in fusion studies is due to the fact that turbulence occurring in the linearly active (unstable) regions of a plasma can travel toward the linearly inactive (stable) regions of the same plasma, where it can modify transport scalings and eventually deteriorate confinement. Spreading of turbulence intensity has been observed in both gyrokinetic simulations \cite{Lin,Yagi,13,14,Migliano} and experiments \cite{15,16,Wu}. It has been shown that turbulence may interact with internal transport barriers and limit their performance \cite{Yagi,Zeng}. In tokamak L mode, turbulence spreading results in global confinement degradation while enhancing ion temperature profile stiffness \cite{Kwon15}. In H mode, turbulence spreading increases the pedestal height and width and hence the energy content of confined plasma \cite{Singh}. Lately, it has been discussed \cite{Nature} that turbulence spreading can mediate the global self-organization of fusion plasma through coupling to other meso-scale phenomena, such as turbulent avalanching \cite{Ben,Politzer,Tokunaga,Canada}, staircasing \cite{DF2010,DF2015,DF2017,Horn2017,Neg,Ashour,Fang,Choi,Van}, and the rise of transport barriers \cite{Zeng,Zonal}. Indirect evidence of turbulence spreading may be obtained from, e.g., the breakdown of gyro-Bohm transport scaling \cite{Lin,Lin02}, the breakdown of Fick's law \cite{Korean,Rev15}, the broadening of scrape-off layer in a tokamak \cite{Wu,Chu}, and the transport shortfall problem \cite{Nature,Short}. Further evidence comes along with cold pulse phenomenology \cite{Gentle,Pulse,Mantica,Mantica_etal} and the observation of pulse polarity inversion \cite{Hariri,Rice13,Rice}. These and other related situations are summarized in Refs. \cite{Singh,Korean}.

From a somewhat more general perspective, turbulence spreading appears to be a key actor behind the transition from weak (quasilinear) to strong (avalanche) transport in magnetic confinement systems \cite{Singh,Hein,Garcia,Zonca05,TJK}. Avalanche transport pertains to a class of nonlocal transport phenomena \cite{Horn2017,Rev15,Korean,Zonca15} for which there is no direct proportionality between fluxes and gradients. A particularly clear example of this is the uphill transport \cite{VM04,VM044,VM06} when fluxes are directed against the gradients, so the effective diffusivity becomes negative \cite{Neg}. Another noted example has referred to the dynamics of very energetic particles (fusion alphas) in a burning plasma \cite{Zonca05,Zonca15,Zonca06,Heid,Zonca_RMF}. There have been some attempts in the literature previously to explain nonlocal transport by implementing a time delay into flux-gradient relations \cite{PoP13,Kosuga}. A further line of inquiry has focussed on emergent behavior of tokamak plasma in proximity to a marginally stable state \cite{Rev15,Diam95}. The phenomenon has been analyzed \cite{Diam95,Newman,Sanchez,Sharma} in the context of self-organized criticality \cite{Bak1,Bak2,Jensen}. The main idea here is that a critical (phase-transitionlike) behavior implies divergence of the spatial correlation length, therefore resulting in the destruction of the local flux-gradient relationship. We hasten to note that the origin of nonlocal transport is far from being completely understood. 

It is worth stressing that turbulence spreading is a property of inhomogeneous turbulence and in that regard lies outside the familiar theories of fully developed turbulence, such as Kolmogorov's K41 theory and other theories alike \cite{Kraichnan,Frisch}.  

A paradigmatic approach to turbulence spreading \cite{Naulin,Itoh,Korean,Gurk05,Zonca06,Gurk06} relies on a conjecture that the transport of turbulence intensity could be described on the basis of a nonlinear diffusion-reaction equation with a combination of sources and sinks$-$which, too, could be made nonlinear$-$so the resulting transport equation is similar in structure to the well-known Fisher-Kolmogorov-Petrovsky-Piskunov equation \cite{Fisher,KPP,Canc}. It has been discussed \cite{Itoh,Gurk05,Gurk06} that the diffusion-reaction model is the minimal analytical model yet accounting for the effects of local linear growth and damping, spatially local nonlinear coupling to dissipation and spatial scattering of turbulence energy induced by nonlinear coupling. 

In the absence of growth or dissipation, the diffusion-reaction model predicts a subdiffusive spreading of an initially localized slug of turbulence in accordance with $\Delta x \propto t^{1/3}$ \cite{Itoh,Gurk05}, where $\Delta x$ is the distance traveled to time $t$. It is understood that the proposed scaling is a mere consequence of the nonlinear diffusion coefficient being a function of turbulence intensity and no more complex than this. In fact, the diffusion-reaction model tacitly assumes quasi-Gaussian fluctuations \cite{Gurk06} and hence finiteness of the mean-squared transport event size. In that regard, the dependence of the diffusion coefficient on fluctuation level is obtained using a weak turbulence closure (i.e., that  the nonlinear spatial scattering is directly proportional to the local turbulence intensity \cite{QL}). In the strong turbulence regime yet consistent with an assumption of local balance between linear growth and damping, a subdiffusive scaling $\Delta x \propto t^{2/5}$ is found \cite{Itoh}, which uses the result of Garbet {\it et al.} \cite{Garbet} that local diffusivity behaves as the square root of local intensity.    

Although the assumptions of Gaussianity and finiteness of second moments can appear restrictive somewhat in terms of a basic theory of turbulent transport, in practice the diffusion-reaction approach proves to be a valuable tool at describing the various aspects of nonlinear spreading dynamics. In fact, by combining the transport equation for turbulence intensity with the equations for temperature, pressure or angular momentum one can characterize a number of phenomena of interest to fusion tasks, such as internal rotation reversals and polarity reversal of cold pulses \cite{Pulse,Hariri}, back reaction of turbulence on plasma profiles \cite{Singh,Guo}, etc. Despite this performance, possible extensions of the diffusion-reaction model to non-Gaussian statistics have also been addressed \cite{Pulse,Gurk06} and related with a class of kinetic equations invoking fractional derivatives \cite{PhD,Report}. As is remarked in Ref. \cite{Gurk06}, the problem with such equations, however, is that they require the corresponding distributions of flight-times and step-sizes \cite{Klafter} as {\it input to the calculation}, instead of predicting them {\it from} the theory.   

In this study, we revise the {\it foundations} of the theory of turbulence spreading. We argue that the propagation of an initially localized slug of turbulence corresponds to a non-Markovian process with a fat-tailed distribution of trapping times. We obtain the parameters of this distribution self-consistently by solving a dynamical problem for a ``particle" interacting with a potential field of the Lennard-Jones type. Concerning the distribution of step-sizes, our result depends on whether the mode-mode interactions occur among three or four waves:  

For three-wave interactions, the distribution of step-sizes is shown to implement a fat tail consistent with the Cauchy (or Lorentz) distribution and therefore suggests the presence of Cauchy flights as an important ingredient to dynamics. A Cauchy flight \cite{Chechkin,Ch2004} is a special case of a L\'evy flight \cite{Klafter,Bouchaud} with the Lorentz distribution of jump lengths. The occurrence of Cauchy-L\'evy flights in three-wave interactions means that the spreading process involves transport events with the divergent second moment \cite{Klafter,Chechkin}, at contrast with the assumptions \cite{Gurk05,Gurk06} of the diffusion-reaction model. 

For four-wave interactions, no L\'evy flights are found instead, indicating that the spreading process has finite variance on all events. 

Either case, the asymptotic ($t\rightarrow+\infty$) spreading appears to be {subdiffusive}, i.e., $\Delta x \propto t^{\nu}$ with $\nu < 1$. More explicitly, we find $\nu = 1/3$ for three-wave interactions and $\nu = 1/4$ for four-wave interactions. These scaling laws are further generalized to a situation according to which the strength of nonlinear interaction depends on width of the field distribution. In a simplest version of this process we find a ballistic scaling $\Delta x \propto t^1$ for three-wave interactions and a diffusive scaling $\Delta x \propto t^{1/2}$ for four-wave interactions. We interpret these behaviors in terms of a wave process with memory.   

The model of turbulence spreading articulated in the present work is very much in the vein of a spreading model considered in Refs. \cite{PRE17,PRE23} with respect to quantum localization of dynamical chaos and the nonlinear Anderson problem \cite{PS,Wang,Flach,Skokos,Iomin,EPL,Fishman,Erez,PRE14,DNC}. A common trait among the two models is the existence of a linearly localized state that is induced by some process: either by spatial disorder$-$as is the case with Anderson localization \cite{And,And+}$-$or by the natural propensity of Fourier components in a tokamak to occur close to their resonance surfaces, as in the model of turbulence spreading proposed in Ref. \cite{Garbet}. When the different components start to overlap, their nonlinear interaction destroys the localized state, giving rise to an unlimited spreading of unstable modes to long distances despite the underlying disorder. The process produces the typical signatures of non-Markovianity and L\'evy flights and is consistent with a description in terms of fractional (or ``strange" \cite{Nat3}) kinetics \cite{Report,Klafter,Sokolov,Rest}.  

The paper is organized as follows. Scaling laws for the asymptotic spreading are obtained first starting from the interaction Hamiltonian (Sec.\,II). We then cast these laws into a kinetic framework using the fractional diffusion equation (Sec.\,III). A discussion of the fractional model is condensed to Sec.\,IV. Section V is concerned with the relaxation dynamics deriving from the spreading process. The keywords here are fractional relaxation equation and the Mittag-Leffler function. Avalanche dynamics is considered in Sec.\,VI, where a fractional modification of the wave equation is advocated. In Sec.\,VII, we consider the possibility that turbulence may tunnel through the regions of regular motion. The focus here is on the shape of the decay function and the phenomena of weak (algebraic) localization. We then obtain the decay function explicitly by solving a bifractional Fokker-Planck equation in a confining potential. Section\,VIII summarizes results.  
   
\section{Scaling laws for turbulence spreading} 

We envisage turbulence as a collection of interacting waves with the dispersion relation $\omega_i = \omega_i (k_i)$, where $\omega_i$ is the frequency of the $i$-th wave, $k_i = |{\bf k}_i|$, and ${\bf k}_i$ is the wave number. The conservation of energy and momentum through the interaction process implies that the interaction cross-section has sharp peak in case of a resonance among the participating wave processes and vanishes otherwise. Respectively for three- and four-wave interactions the conditions for a resonance read
\begin{equation}
\omega_k = \omega_{k_1} + \omega_{k_2},\ \ \ {\bf k} = {\bf k}_1 + {\bf k}_2,
\label{R3} % Eq.~(\ref{R3})
\end{equation} 
and 
\begin{equation}
\omega_{k_1} + \omega_{k_2} = \omega_{k_3} + \omega_{k_4},\ \ \ {\bf k}_1 + {\bf k}_2 = {\bf k}_3 + {\bf k}_4,
\label{R4a} % Eq.~(\ref{R4a})
\end{equation} 
\begin{equation}
\omega_{k_1} = \omega_{k_2} + \omega_{k_3} + \omega_{k_4},\ \ \ {\bf k}_1 = {\bf k}_2 + {\bf k}_3 + {\bf k}_4.
\label{R4b} % Eq.~(\ref{R4b})
\end{equation}
Whether these conditions are actually matched for some frequencies and wave numbers is decided by the concrete $\omega_i = \omega_i (k_i)$ dependence (i.e., by the specific instability at play: see Sec.\,VIII) \cite{QL,Sagdeev}. 

In order to predict the scaling laws for asymptotic spreading, one needs to know how the different resonances are folded in the ambient phase space and if there is an overlap between them. As Chirikov \cite{Chirikov} realized, any overlap between resonances will introduce a local instability into the dynamics, which renders a system non-integrable. This is because the overlapping modes may share their resonances and by doing so occasionally switch from one resonance to another, thus giving rise to a stochastic motion process in phase space. Applying this argument to the spreading problem one concludes that turbulence spreading to long distances is limited to a set of overlapping resonances stretching to infinity (in practice, to outer boundaries of the finite-size system). 

Perhaps the most straightforward situation permitting a connected escape path to infinity is when the overlapping resonances densely fill the ambient space. This is fully developed chaos \cite{Report,Sagdeev}, a classic model of chaotic dynamics leading to the familiar Fokker-Planck equation \cite{ZaslavskyUFN,ChirikovUFN}. A problem with fully developed chaos, however, is that it corresponds to a space-filling turbulence, while the phenomenon of turbulence spreading would imply that the turbulence is {\it inhomogeneous}. Another problem \cite{EPL,PRE14} is that fully developed chaos requires a divergent energy reservoir in systems with a large number of interacting degrees of freedom. On the contrary, in most situations of practical significance one inquires into spreading of a small slug of turbulence that is not supposed to visit the entire phase space$-$such as, for instance, in the models of cold pulse propagation \cite{Pulse,Hariri} or turbulence penetration from the plasma edge into the scrape-off layer of a tokamak \cite{Wu,Chu}. In the nonlinear Anderson problem, an assumption that chaos is strong and fills the ambient space leads to an incorrect asymptotic scaling law that is not confirmed through simulations \cite{PS,Flach,Skokos,Ivan,Senyange}. Based on this evidence, we disregard the involvement of fully developed chaos in the present theory. 

In practical terms, we have to admit that the phase space of a dynamical system with turbulence spreading could be actually very complex and strongly shaped, and include both the areas of strong chaos (so-called stochastic sea \cite{Report}) and islands of nearly regular dynamics. As is noted by Zaslavsky \cite{Report}, the presence of islands in phase space means automatically a kind of non-ergodicity, since no trajectory from the stochastic sea can ever penetrate the islands, and vice versa. This no-entry rule appears to be a ``trouble" point of chaotic dynamics as it results in stickiness to boundaries, dynamical traps and other unpleasant features alike \cite{Report,Trap,Pseudo,Edel}.     

Another aspect worth noting concerns the involvement of low frequencies (in the sense of Refs. \cite{Misguich,JJR,Horton,PRE01,PRE09}) into the asymptotic spreading. Indeed, such frequencies, even if quiescent from the outset, will be naturally excited through nonlinear coupling processes as the turbulent field spreads to large spatial scales. This opens up the doorway towards the resonant process    
\begin{equation}
\omega = \omega_{k_j} + \omega_{-k_j} \simeq 0,\ \ \ {\bf k}_j + ({-\bf k}_j) \simeq 0,
\label{R0} % Eq.~(\ref{R0})
\end{equation} 
where three-wave interactions have been assumed. The set of equations~(\ref{R0}) is obtained from the resonance condition in Eq.~(\ref{R3}) by letting $\omega_k = \omega$ and $\omega \rightarrow 0$. If the turbulent field is characterized by a broad spectrum of frequencies and wave vectors$-$as most theories of wave turbulence would imply \cite{QL}$-$then the overlap of a large number of resonances satisfying~(\ref{R0}) will generate a frequency spread around $\omega$ of the order of $\Delta \omega \propto {\sqrt{|\omega|}}$, if the law of large numbers \cite{LNL} holds. This frequency spread is none other than the width of the stochastic layer in the limit $\omega\rightarrow 0$ \cite{Sagdeev,ZaslavskyUFN}. On the other hand, the distance between neighboring resonances in the frequency domain behaves as $\delta\omega\propto\omega$ and for $\omega\rightarrow 0$ will be by far smaller than $\Delta \omega \propto {\sqrt{|\omega|}}$. That means that the zero-frequency resonance is {\it always} surrounded by a layer of stochastic dynamics, provided just that the field is spread over a sufficiently large number of states. Note that the number density of overlapping resonances inside the layer diverges as $\propto 1/\omega\rightarrow\infty$.  

In tokamak applications, the zero-frequency resonance in Eq.~(\ref{R0}) is customarily associated with zonal flows. In a dedicated theory \cite{Zonal,Zonal06} it is shown that zonal flows are driven exclusively by nonlinear interactions which transfer energy from the finite-$j$ drift waves to the $j=0$ flow. This energy-transfer process directed towards large spatial scales is generic to 2D turbulence, both in plasmas \cite{JJR,Horton,Huld,PLAN} and fluids \cite{Kraichnan,Frisch}. If one assumes, following the wisdom of self-organized criticality \cite{Bak1,Jensen}, that the low frequencies in Eq.~(\ref{R0}) are excited dynamically through turbulent spreading of drift waves, one infers that spreading to long distances results in$-$and via the complexity synergy in fusion plasma also results from$-$the occurrence of zonal flows. Such synergetic cooperation between zonal flows and turbulence spreading has been addressed in Refs. \cite{Korean,Singh,Nature,Short}, where further insight into turbulence self-regulation can be found. 

With these premises, we are in position to obtain the asymptotic scaling laws for turbulence spreading. 
 
\subsection{Three-wave interactions}

The Hamiltonian of three-wave interaction on a lattice reads \cite{Sagdeev}
\begin{equation}
H = H_{0} + H_{\rm int}, \ \ \ H_0 = \frac{1}{2}\sum_k \omega_k \sigma^*_k \sigma_k,
\label{H0} % Eq.~(\ref{H0})
\end{equation}
\begin{equation}
H_{\rm int} = \frac{1}{3}\sum_{k, k_1, k_2} V_{-k, k_1, k_2} \sigma^*_{k} \sigma_{k_1} \sigma_{k_2}\delta_{-k+k_1+k_2},
\label{3W} % Eq.~(\ref{3W})
\end{equation}
where $H_0$ is the Hamiltonian of noninteracting waves, $H_{\rm int}$ is the interaction Hamiltonian, $\sigma_k = \sigma_k (t)$ are complex amplitudes which represent a wave process with frequency $\omega_k$ and wave number ${k}$ and which may depend on time $t$ in general, $\sigma_{-k} = \sigma_k^*$ owing to the time symmetry $t\leftrightarrow -t$, the asterisk denotes complex conjugate, $k=0,\pm 1,\pm 2,\dots$ on a discrete lattice, the lattice step has been set to 1, $V_{-k, k_1, k_2}$ are complex coefficients which characterize the cross-section of the interaction process $k = k_1 + k_2$, and $\delta_{-k+k_1+k_2}$ is the Kroneker delta which accounts for resonant character of the interactions. We consider a 1D lattice for simplicity (see, however, a remark in Sec. III\,D). The 1D problem will directly apply to the case of turbulence spreading in one preferred direction such as, e.g., the perpendicular direction to the magnetic field in magnetically confined tokamak plasmas, at no contradiction with the fact that the actual system may be 2D or 3D.    

Given the Hamiltonian in Eqs.~(\ref{H0}) and~(\ref{3W}), one applies the canonical equations 
\begin{equation}
\dot{\sigma}_k = i\frac{\partial H}{\partial\sigma_k^*}, \ \ \  \dot{\sigma}_k^* = -i\frac{\partial H}{\partial\sigma_k}
\label{Can} % Eq.~(\ref{Can})
\end{equation}
to obtain the equation of motion for the complex amplitude ${\sigma_k}$, i.e., 
\begin{equation}
\dot{\sigma}_k = i\omega_k\sigma_k + iV^*\sigma_{k_1} \sigma_{k_2}.
\label{DEM} % Eq.~(\ref{DEM})
\end{equation}
The equations for $\sigma_{k_1}$ and $\sigma_{k_2}$ are obtained either directly from the Hamiltonian $H = H_0 + H_{\rm int}$ by applying the canonical equations, or by switching indexes in the equation of motion~(\ref{DEM}) on account of the resonance condition $k=k_1 + k_2$ and the general rule $\sigma_{-k} = \sigma_k^*$. The end result is 
\begin{equation}
\dot{\sigma}_{k_1} = i\omega_{k_1}\sigma_{k_1} - iV\sigma_{k} \sigma_{k_2}^*,
\label{DEM1} % Eq.~(\ref{DEM1})
\end{equation}
\begin{equation}
\dot{\sigma}_{k_2} = i\omega_{k_2}\sigma_{k_2} - iV\sigma_{k} \sigma_{k_1}^*,
\label{DEM2} % Eq.~(\ref{DEM2})
\end{equation}
where $V \equiv V_{k, -k_1, -k_2}$, and we have used a dot to denote time differentiation. 

If the field is spread across $\Delta n \gg 1$ states, then the conservation of the total probability 
\begin{equation}
\sum_{n=0}^{\Delta n} |\sigma_n|^2 \simeq \int_0^{\Delta n} |\sigma_n|^2 d n = 1
\label{TP} % Eq.~(\ref{TP})
\end{equation}
would imply that the density of the probability is small and is inversely proportional to $\Delta n$, i.e., $|\sigma_n|^2 \simeq 1/\Delta n$. It is understood that the spreading of the wave field in phase space is due to the nonlinear interactions among the unstable modes. For small amplitudes, which we consider, these interactions are reduced to triad interactions in the leading order (provided just that such interactions are permitted by the dispersion relation). The rate of field spreading can therefore be obtained from the dynamical equations~(\ref{DEM})$-$(\ref{DEM2}) to give
\begin{equation}
R\simeq |\dot{\sigma}_n|^2 \simeq |V|^2|\sigma_n|^4\simeq |V|^2/(\Delta n)^2.
\label{Rate} % Eq.~(\ref{Rate})
\end{equation}
On the other hand, the rate $R$ can be written on account of Fermi's golden rule \cite{Golden} as $R \simeq d\Delta n/ dt$ consistently with the resonant character of interactions. Combining with Eq.~(\ref{Rate}), one obtains
\begin{equation}
d\Delta n / dt = \zeta |V|^2/(\Delta n)^2, 
\label{Rate2} % Eq.~(\ref{Rate2})
\end{equation}
where $\zeta$ is a coefficient providing the equity sign to the above equation. Integrating over time in Eq.~(\ref{Rate2}), one gets $(\Delta n)^3 = 3\zeta |V|^2 t$, from which a subdiffusive spreading law  
\begin{equation}
(\Delta n)^2 = (3\zeta)^{2/3} |V|^{4/3} t^{2/3}
\label{SpL} % Eq.~(\ref{SpL})
\end{equation}
can be deduced for $t\rightarrow+\infty$.

At first glance, the spreading law in Eq.~(\ref{SpL}) conforms to the scaling $\Delta x \propto t^{1/3}$ obtained in Refs. \cite{Itoh,Gurk05} from the diffusion-reaction model. There is a subtlety, however, and this is seen from the fact that the ``diffusion coefficient" in Eq.~(\ref{SpL}) is proportional to the power $2/3$ of the cross-section parameter $|V|^2$, while in the model of G\"urcan {\it et al.} \cite{Gurk05} it is proportional to intensity of the turbulent field. The deviation occurs, because in our model we do {\it not} really assume that the spreading process has diffusive nature, nor do we involve any form of quasilinear theory to calculate the diffusion constant. Instead, our analysis has focussed on the resonant character of mode-mode coupling and the fact that this coupling is triad-like. No Gaussianity of fluctuations has been invoked in contrast to Refs. \cite{Itoh,Gurk05,Gurk06}. We strengthen this argument in Sec.\,III, where a kinetic equation for turbulence spreading is derived.   

\subsection{Four-wave interactions} 

Mathematically, this case is similar to the three-wave interaction case considered previously. For the resonant four-wave interactions, the interaction Hamiltonian takes the form
\begin{equation}
H_{\rm int} = \frac{1}{4}\sum_{k, k_1, k_2, k_3} V_{-k, k_1, k_2, k_3} \sigma^*_{k}\sigma_{k_1}\sigma_{k_2}\sigma_{k_3}\delta_{-k+k_1+k_2+k_3},
\label{4W} % Eq.~(\ref{4W})
\end{equation}
from which the following equation of motion for the complex amplitude ${\sigma_k}$ may be deduced:
\begin{equation}
\dot{\sigma}_k = i\omega_k\sigma_k + iV^*\sigma_{k_1} \sigma_{k_2} \sigma_{k_3},
\label{DEM4} % Eq.~(\ref{DEM4})
\end{equation}
where $V \equiv V_{k, -k_1, -k_2, -k_3}$. Switching the indexes in Eq.~(\ref{DEM4}), and remembering that $\sigma_{-k} = \sigma_k^*$, one gets the dynamical equation for ${\sigma_{k_1}}$, i.e., 
\begin{equation}
\dot{\sigma}_{k_1} = i\omega_{k_1}\sigma_{k_1} - iV\sigma_{k} \sigma_{k_2}^* \sigma_{k_3}^*,
\label{DEM4-1} % Eq.~(\ref{DEM4-1})
\end{equation}
and similarly for ${\sigma_{k_2}}$ and ${\sigma_{k_3}}$.

The rate of field spreading is obtained as 
\begin{equation}
R\simeq |\dot{\sigma}_n|^2 \simeq |V|^2|\sigma_n|^6\simeq |V|^2/(\Delta n)^3,
\label{Rate-ext} % Eq.~(\ref{Rate-ext})
\end{equation}
where the conservation of the total probability through the spreading process has been considered, i.e., $|\sigma_n|^2 \simeq 1/\Delta n$. Writing the rate $R$ as $R \simeq d\Delta n/ dt$ in accordance with Fermi's golden rule, one gets
\begin{equation}
d\Delta n / dt = \zeta |V|^2/(\Delta n)^3, 
\label{Rate2-ext} % Eq.~(\ref{Rate2-ext})
\end{equation}
from which $(\Delta n)^4 = 4\zeta |V|^2 t$. This corresponds to a subdiffusive spreading for $t\rightarrow+\infty$, i.e., 
\begin{equation}
(\Delta n)^2 = (4\zeta)^{1/2} |V| t^{1/2},
\label{SpL-ext} % Eq.~(\ref{SpL-ext})
\end{equation}
where $\zeta$ is a coefficient of the spreading process. 

The spreading law in Eq.~(\ref{SpL-ext}) is actually a familiar one. It has been obtained both theoretically \cite{PRE17,EPL23,PRE23} and numerically \cite{Senyange} for the spreading of an initially localized wave packet in nonlinear Schr\"odinger lattices with disorder. Also it has been obtained for quantum chaotic subdiffusion in random potentials by Ivanchenko {\it et al.} \cite{Ivan} in the framework of the Hubbard model. Note that the ``diffusion coefficient" pertaining to the spreading process in Eq.~(\ref{SpL-ext}) is proportional to the square root of $|V|^2$. 

\section{Bifractional kinetic equation} 

We begin with showing that the subdiffusive scaling laws in Eqs.~(\ref{SpL}) and~(\ref{SpL-ext}) correspond to a non-Markovian spreading process with a waiting time statistics. For this, let us unify these scalings first with the aid of   
\begin{equation}
d\Delta n / dt = A /(\Delta n)^{2s+1}, 
\label{RR} % Eq.~(\ref{RR})
\end{equation}
where the switcher $s$ takes the value $s=1$ for four-wave interactions and the value $s=1/2$ for three-wave interactions, and we have denoted $A = \zeta |V|^2$ for simplicity. Integrating over time in Eq.~(\ref{RR}), one gets $(\Delta n)^{2s+2} = (2s+2)At$, from which
\begin{equation}
(\Delta n)^2 = [(2s + 2) A]^{1/(s+1)}t^{1/(s+1)}
\label{Sub} % Eq.~(\ref{Sub})
\end{equation}
for $t\rightarrow+\infty$. Equation~(\ref{Sub}) interpolates between the two scaling laws above by allowing the switcher $s$ to vary from $s=1/2$ in Eq.~(\ref{SpL}) to $s=1$ in Eq.~(\ref{SpL-ext}). It is noted that the behavior of the ``diffusion coefficient" with the cross-section of the interaction process is represented by the power $1/(s+1)$ of $|V|^2$.  

Differentiating both sides of Eq.~(\ref{RR}) with respect to time and eliminating the remaining $d\Delta n / dt$ with the aid of the same Eq.~(\ref{RR}), one gets 
\begin{equation}
\frac{d^2 \Delta n}{dt^2}= -\frac{(2s + 1)A^2}{(\Delta n)^{4s + 3}}.
\label{Grad+} % Eq.~(\ref{Grad+})
\end{equation}
Using a gradient form on the right-hand side of Eq.~(\ref{Grad+}), one obtains
\begin{equation}
\frac{d^2}{dt^2}\Delta n = - \frac{d}{d \Delta n} \left(- \frac{A^2 / 2}{(\Delta n)^{4s + 2}}\right).
\label{Grad} % Eq.~(\ref{Grad})
\end{equation}
Equation~(\ref{Grad}) is none other than a Newtonian equation of motion along the coordinate $\Delta n$ in the potential field 
\begin{equation}
W (\Delta n) = - \frac{A^2 / 2}{(\Delta n)^{4s + 2}}.
\label{Poten} % Eq.~(\ref{Poten})
\end{equation}
For $s = 1$, the potential function in Eq.~(\ref{Poten}) becomes $W(\Delta n) = - (A^2 / 2)/ (\Delta n)^6$ and is immediately recognized as the attractive part of the celebrated Lennard-Jones potential \cite{Lennard} known from molecular physics. 

Given the attractive character of $W (\Delta n)$, one might arguably propose that the newly excited modes will tend to form clusters (``molecules") in phase space, where they will be effectively trapped \cite{Iomin,PRE17} due to their nonlinear coupling. It is due to these attractive ``forces" acting among the components of the wave field that the asymptotic transport deviates from normal diffusion in the limit $t\rightarrow+\infty$. It is understood that the origin of these forces is rooted in the non-ergodic nature of the spreading process and the presence of multiple islands of regular dynamics. Indeed, the effect of islands is such as to squeeze the stochastic motions into narrow layers {\it between} the islands, impeding exponentially fast separation of dynamical trajectories for $t\rightarrow+\infty$. As a result, nearby trajectories stick together for asymptotically long times, which is equivalent to an attractive interaction among them. Such ``stickiness" of dynamical trajectories for $t\rightarrow+\infty$ implies that the Lyapunov exponents vanish in the thermodynamic limit, despite that the dynamics are random. This type of behavior is a defining feature of weak chaos \cite{PhD,Report,Pseudo,Edel}. 

One sees that the asymptotic ($t\rightarrow+\infty$) spreading corresponds to a {\it weakly} chaotic process with stickiness, and not really to strong chaos with a space-filling stochastic sea, contrary to a common belief. We should stress that the hypothesis of weak chaos excludes that the asymptotic spreading could be a Gaussian random walk.         

An alternative way to understand clustering is to associate it with spontaneous synchronization of the different modes. The main idea here \cite{Winfree,Strogatz} is that a set of coupled nonlinear oscillators could exhibit the temporal analog of a phase transition. When the spread of natural frequencies is large compared to the coupling frequency, the system behaves incoherently, with the nonlinear interaction term acting as a white noise \cite{EPL,Erez}. But when the spread goes below a certain critical limit, then a small cluster of oscillators spontaneously locks into a synchronized state \cite{Strogatz}. In non-ergodic systems, a limitation on frequency spread is naturally induced by the islands of regular dynamics (because islands favor stickiness of trajectories in the long run). If, as we assume, there exist islands of arbitrarily large size, then there will {\it always} be areas of phase space with too narrow a frequency spread to permit clustering of trajectories. The occurrence of clusters results in a non-Markovian spreading for $t\rightarrow+\infty$, with a distribution of trapping, or exit, times, as we now proceed to show.  

Multiplying both sides of Eq.~(\ref{Grad}) by the velocity $d \Delta n /dt$ and integrating the ensuing differential equation with respect to time, after a simple algebra one obtains
\begin{equation}
\frac{1}{2}\left(\frac{d}{dt} \Delta n\right)^2 - \frac{A^2 / 2}{(\Delta n)^{4s + 2}} = \Delta E,
\label{Ener} % Eq.~(\ref{Ener})
\end{equation}
where the first term on the left-hand side has the sense of the kinetic energy of a ``particle" of unit mass moving along the coordinate $\Delta n$ and the second term is its potential energy. It is shown using Eq.~(\ref{RR}) that the kinetic energy in Eq.~(\ref{Ener}) compensates for the potential energy {exactly}, that is, the full energy in Eq.~(\ref{Ener}) is zero, {i.e.}, $\Delta E = 0$. More so, both the negative potential energy $W (\Delta n) \sim - A^2 / 2(\Delta n)^{4s + 2}$ and the positive kinetic energy $\frac{1}{2}(d \Delta n /dt)^2 \sim A^2 / 2(\Delta n)^{2(2s + 1)}$ vanish while spreading. Both will decay as the $(4s + 2)$th power of the number of states and the ratio between them will {\it not} depend on width of the field distribution. 

The full energy being equal to zero implies that the particle in Eq.~(\ref{Ener}) is sitting on the separatrix $\Delta E = 0$. The separatrix character means, in its turn, that the dynamics remain unstable on all time scales up to $t\rightarrow+\infty$. This is due to the very peculiar nature of separatrix transport \cite{Sagdeev,ZaslavskyUFN}, where tiny perturbations caused by, e.g., a small white noise or imprecision in the initial conditions can drastically change the type of phase-space trajectory. The observation is especially relevant for separatrix dynamics in large systems \cite{ChV}. Indeed, the fact that such systems can extend to arbitrarily long spatial scales implies that the density of overlapping resonances in vicinity of the separatrix boosts as $\propto 1/\omega$ for $\omega\rightarrow 0$ (Sec.\,II). This produces kind of a fluctuation background or thermal ``bath" enveloping the separatrix, which imposes a spread on low-frequency interactions of the order of $\Delta \omega \propto {\sqrt{|\omega|}}$. Naturally the bath influences the possibility for dynamical trajectories to freeze into clusters, but it also limits the time the trajectories can spend within clusters. To evaluate these features, one needs to assess how the Lennard-Jones potential in Eq.~(\ref{Poten}) weighs against the strength of fluctuations in the stochastic layer.  

\subsection{Waiting-time distribution}

To perform this task, let us assume, following Refs. \cite{PRE17,PRE23}, that the fluctuation background around the separatrix is characterized by the thermodynamic temperature $T$. That is, the value of $T$ measures how intense are the random motions in the stochastic layer. Crudely put, $T$ characterizes the density of overlapping resonances inside the layer, while the exact relationship could be a matter of the equation of state. Then the probability for a given mode to quit the cluster after it has traveled $\Delta n$ sites on it is given by the Boltzmann factor 
\begin{equation}
p (\Delta n) = \exp [W (\Delta n) / T],
\label{BF} % Eq.~(\ref{BF})
\end{equation}
where we have set the Boltzmann constant to unity. Substituting $W (\Delta n)$ from the potential function in Eq.~(\ref{Poten}), one finds
\begin{equation}
p (\Delta n) = \exp [- A^2 / 2 T (\Delta n)^{4s + 2}].
\label{Escape} % Eq.~(\ref{Escape})
\end{equation}
Taylor expanding the exponential function for $\Delta n \gg 1$, one gets 
\begin{equation}
p (\Delta n) \simeq 1 - A^2 / 2 T (\Delta n)^{4s + 2}.
\label{Expand} % Eq.~(\ref{Expand})
\end{equation}
The probability to remain (survive) on the cluster after $\Delta n$ space steps is $p^{\prime} (\Delta n) = 1-p(\Delta n)$, yielding 
\begin{equation}
p^{\prime} (\Delta n) \simeq A^2 / 2 T (\Delta n)^{4s + 2}.
\label{EscapePr} % Eq.~(\ref{EscapePr})
\end{equation}
Eliminating $\Delta n$ with the aid of Eq.~(\ref{Sub}), one obtains the probability to survive on the cluster after $\Delta t$ time steps  
\begin{equation}
p^{\prime} (\Delta t) \propto (\Delta t)^{-(2s+1) / (s+1)},
\label{Survive} % Eq.~(\ref{Survive})
\end{equation}
leading to a waiting-time distribution \cite{Klafter,Sokolov} 
\begin{equation}
\chi_\alpha (\Delta t) \propto (\Delta t)^{-(1+\alpha)}
\label{WT} % Eq.~(\ref{WT})
\end{equation}
with $\alpha = s/(s+1) < 1$. We associate the distribution in Eq.~(\ref{WT}) with the binding effect of finite clusters \cite{Iomin,PRE17}. Note that the integral 
\begin{equation}
\int_{\sim 1}^{\tau} \Delta t \chi_\alpha (\Delta t) d\Delta t \sim \int_{\sim 1}^{\tau} (\Delta t)^{-\alpha} d\Delta t \sim \tau^{1-\alpha} \rightarrow+\infty
\label{Div} % Eq.~(\ref{Div})
\end{equation}
diverges for $\tau\rightarrow+\infty$, implying that the mean waiting time is infinite for all $\alpha < 1$. Specifically, we find $\alpha = 1/3$ for three-wave interactions ($s=1/2$) and $\alpha = 1/2$ for four-wave interactions ($s=1$). 

The fact that the mean waiting time diverges for $\alpha < 1$ is explained by the presence of multiple islands of regular dynamics in phase space, imposing stickiness and weak mixing \cite{Report,Pseudo} in the limit $t\rightarrow+\infty$. 

\subsection{Non-Markovian diffusion equation}

Let us now obtain a kinetic equation for the asymptotic spreading. For this, we adopt the theoretical scheme of continuous time random walks (CTRW) \cite{Bouchaud,CTRW1,CTRW2}, according to which the transport occurs as a result of random-walk jumps in phase space, with a distribution of waiting times between consecutive steps of the motion. Combining~(\ref{WT}) with a simplifying assumption (to be revisited below) that there is a characteristic jump length of the random process, one obtains a non-Markovian generalization of the diffusion equation \cite{Klafter,Sokolov,Rest}, i.e., 
\begin{equation}
\frac{\partial}{\partial t} f (n, t) =\,_{0}{D}_t^{1-\alpha}\Big(K_{\alpha} \frac{\partial^{2}}{\partial n^{2}} f (n, t)\Big).
\label{FDEL-L} % Eq.~(\ref{FDEL-L})
\end{equation}
Here, $f = f (n, t)$ is the probability density to find the random walker at time $t$ at the distance $n$ away from the origin, $K_{\alpha}$ is the transport coefficient of the dimension $[K_{\alpha}]=$~cm$^{2}\,\cdot\,$sec$^{-\alpha}$, and 
\begin{equation}
{_0}{D}_t^{1-\alpha} f (n, t) = \frac{1}{\Gamma (\alpha)}\frac{\partial}{\partial t}\int _{0}^{t} \frac{f (n, t^{\prime})}{(t - t^{\prime})^{1-\alpha}}dt^\prime \label{R-L} % Eq.~(\ref{R-L})
\end{equation}
is the Riemann-Liouville fractional derivative \cite{Podlubny,Samko} of order $1-\alpha$. The latter type of ``derivative" is in fact an integro-differential operator with a power-law memory kernel. It accounts for multi-scale trappings of unstable modes within phase-space clusters \cite{Iomin,PRE17} consistently with the waiting-time distribution in Eq.~(\ref{WT}). Note that we directly associate the non-Markovian character of Eq.~(\ref{FDEL-L}) with the binding effect of the Lennard-Jones potential in Eq.~(\ref{Poten}) and the fact that the mean waiting time diverges for $\tau\rightarrow+\infty$. Based on kinetic Eq.~(\ref{FDEL-L}), one finds the asymptotic ($t\rightarrow+\infty$) mean-square displacement of the random walker to be    
\begin{equation}
\langle (\Delta n)^2 (t) \rangle \propto t^{\alpha},
\label{MSD} % Eq.~(\ref{MSD})
\end{equation} 
where $\alpha = s/(s+1)$. Because $\alpha < 1$ for any finite $s \geq 0$, the spreading process is subdiffusive. 

\subsection{Inconsistency for $s < 1$}

If one compares the scaling law in Eq.~(\ref{MSD}) with that in Eq.~(\ref{Sub}), one sees that both are consistent {\it only} if $s=1$, while for $s < 1$ there is a discrepancy to be repaired in a way. Indeed, the scaling in Eq.~(\ref{MSD}) suggests $\langle (\Delta n)^2 (t) \rangle \propto t^{s/(s+1)}$, whereas the structure of interactions dictates $(\Delta n)^2 (t) \propto t^{1/(s+1)}$. One concludes that the non-Markovian model in Eq.~(\ref{FDEL-L}) could be a correct transport model in case of four-wave interactions, while with three-wave interactions it is not satisfactory yet.    

The observed discrepancy is resolved by assuming \cite{PRE19,EPL23,PRE23} that the waiting-time statistics in Eq.~(\ref{WT}) competes with a fat-tailed distribution of step-sizes of the random walk  
\begin{equation}
\chi_\mu (|\Delta n|) \propto |\Delta n|^{-(1+\mu)},
\label{LW} % Eq.~(\ref{LW})
\end{equation}    
where $0 < \mu < 2$ is a power exponent. We associate the step-size distribution in Eq.~(\ref{LW}) with multi-scale jumps between different clusters.    

\subsection{Non-Markovian equation with jumps}

The inclusion of jumps leads to a more general equation \cite{Klafter,Chechkin} for the probability density $f = f (n, t)$, i.e.,   
\begin{equation}
\frac{\partial}{\partial t} f (n, t) =\,_{0}{D}_t^{1-\alpha}\Big(K^{\mu}_{\alpha} \frac{\partial^{\mu}}{\partial |n|^{\mu}} f (n, t)\Big),
\label{FDEL} % Eq.~(\ref{FDEL})
\end{equation}
where in place of $\partial^2/\partial n^2$ one encounters the Riesz fractional derivative of order $\mu$, that is \cite{Podlubny,Samko} 
\begin{equation}
\frac{\partial^\mu}{\partial |n|^\mu} f (n, t) = \frac{1}{\Gamma_\mu}\frac{\partial^2}{\partial n^2} \int_{-\infty}^{+\infty}\frac{f (n^\prime, t)}{|n-n^\prime|^{\mu - 1}} dn^\prime.
\label{Def+} % Eq.~(\ref{Def+})
\end{equation} 
It is understood that the Riesz derivative differs from the analogue Riemann-Lioville derivative in Eq.~(\ref{R-L}) in that it applies to the position coordinate $n$ in infinite space: $-\infty < n < +\infty$. In the above, $\Gamma_\mu = -2\cos(\pi\mu/2)\Gamma(2-\mu)$ is a normalization parameter, which is introduced to ensure a smooth crossover to $\partial^2/\partial n^2$ for $\mu\rightarrow 2$, $K^{\mu}_{\alpha}$ is the generalized transport coefficient of the dimension $[K^{\mu}_{\alpha}]=$~cm$^{\mu}\,\cdot\,$sec$^{-\alpha}$, and we have tacitly assumed that the exponent $\mu$ lies within $1 < \mu < 2$. The latter assumption guarantees that the step-size distribution in Eq.~(\ref{LW}) belongs to a class of L\'evy stable distributions \cite{Bouchaud,Gnedenko}. The interval $0 < \mu < 1$, although similar, is not considered here. As is well known \cite{Klafter,Chechkin,Rest}, the Riesz derivative in kinetic Eq.~(\ref{FDEL}) generates L\'evy flights. 

Note that the Riesz derivative incorporates nonlocal features to the transport [via the improper integration over $n^\prime$ on the right-hand side of Eq.~(\ref{Def+})]. Note, also, that the properties of non-Markovianity and nonlocality are attributed to two different operators$-$Eqs.~(\ref{R-L}) and~(\ref{Def+})$-$with the integration along time and position coordinate, respectively. 

Another technicality worth noting is that we write all transport equations in 1D. This 1D transport model is indeed often what one is looking for in the applications \cite{Naulin,Hariri,Pulse,Zonca05,Zonca06} and therefore is important for interpretation of results. A generalization to higher embedding dimensions is obtained straightforwardly by upgrading the Riesz fractional derivative to the Riesz/Weyl operator \cite{Klafter,Rest}. Mathematically, this step is actually very well known in fractional calculus \cite{PhD,Report}.

If $\mu\rightarrow 2$, then the Riesz derivative in Eq.~(\ref{Def+}) is none other that the familiar second-order spatial derivative, i.e., $\lim_{\mu\rightarrow 2} {\partial^{\mu}}/\partial {|n|^{\mu}} = {\partial^{2}}/\partial {n^{2}}$. To this end, nonlocality is lost giving rise to a local kinetic process in phase space. For $\mu\rightarrow 1$, the exact analytical representation of ${\partial^{\mu}}/\partial {|n|^{\mu}}$ is obtained via a reduction of Eq.~(\ref{Def+}) to the Hilbert transform operator \cite{Chechkin,Mainardi}, yielding 
\begin{equation}
\frac{\partial^\mu}{\partial |n|^\mu} f (n, t)  = - \frac{1}{\pi} \frac{\partial}{\partial n} \int_{-\infty}^{+\infty}\frac{f (n^\prime, t)}{n-n^\prime} dn^\prime, \ \ \ \mu=1.
\label{Hilbert} % Eq.~(\ref{Hilbert})
\end{equation} 
It is understood that the Hilbert transform operator~(\ref{Hilbert}) produces Cauchy flights \cite{Chechkin,Ch2004}$-$a specific variant of L\'evy flights with the Lorentz distribution of jump lengths.  

Using kinetic Eq.~(\ref{FDEL}), one obtains the fractional moments \cite{Klafter} of the $f = f (n, t)$ distribution, from which the scaling of the pseudo-mean square displacement may be deduced for $t\rightarrow+\infty$, leading to 
\begin{equation}
\langle (\Delta n)^2 (t) \rangle = \lim_{\delta\rightarrow 2}\,(\Delta n)^\delta \propto t^{2\alpha/\mu}.
\label{Pseudo} % Eq.~(\ref{Pseudo})
\end{equation} 
An exact calculation of the fractional moments of $f (n, t)$ invokes the formalism of the Fox $H$-functions \cite{Podlubny,Samko} and is articulated in Refs. \cite{Klafter,Rest,Chechkin}. Comparing to Eq.~(\ref{Sub}), one infers 
\begin{equation}
2\alpha/\mu = 1/(1+s),
\label{Inf} % Eq.~(\ref{Inf})
\end{equation} 
from which 
\begin{equation}
\mu = 2s,
\label{Inf+} % Eq.~(\ref{Inf+})
\end{equation} 
where the general relation $\alpha = s/(s+1)$ stemming from Eq.~(\ref{WT}) has been considered. 

For $s=1$, Eq.~(\ref{Inf+}) predicts $\mu = 2$ consistently with the kinetic Eq.~(\ref{FDEL-L}). That means that four-wave interactions produce a non-Markovian transport pattern without nonlocality. Indeed, the transport model in Eq.~(\ref{FDEL-L}) is a differential one with respect to the spatial coordinate $n$. Yet, it includes a memory into the spreading process due to the presence of the Riemann-Liouville fractional derivative on the right-hand side of Eq.~(\ref{FDEL-L}). 

The situation changes drastically for three-wave interactions ($s=1/2$) for which one finds a very nontrivial value $\mu = 1$ instead. The latter value corresponds to the Hilbert transform operator~(\ref{Hilbert}). It strongly suggests the presence of Cauchy-L\'evy flights through the dynamics of unstable modes. The corresponding transport equation is qualitatively different from the one in Eq.~(\ref{FDEL-L}) in that it is explicitly nonlocal (involves improper integration over $n^\prime$ with a power-law kernel of the Pareto-L\'evy type). On account of the kinetic model in Eq.~(\ref{FDEL}) the resulting transport equation reads
\begin{equation}
\frac{\partial}{\partial t} f (n, t) = -\frac{1}{\pi}\,_{0}{D}_t^{1-\alpha} \Big(K^{1}_{\alpha}\frac{\partial}{\partial n} \int_{-\infty}^{+\infty}\frac{f (n^\prime, t)}{n-n^\prime} dn^\prime\Big),
\label{Hil} % Eq.~(\ref{Hil})
\end{equation}
where the transform operator in Eq.~(\ref{Hilbert}) has been used. One sees that a spreading process resulting from three-wave interactions is actually a complex one. It involves if only non-Markovianity due to the Riemann-Liouville fractional derivative on the right-hand side of Eq.~(\ref{Hil}) as well as an explicit nonlocality with respect to the position coordinate $n$ (owing to the Hilbert transform operator). The subdiffusion in Eq.~(\ref{Pseudo}) is therefore a result of a competition between the non-Markovian and nonlocal features under the general condition in Eq.~(\ref{Inf}). 

\subsection{Brief summary}

Our findings so far can be summarized as follows. Turbulence spreading has complex microscopic organization, it is nondiffusive and intermittent in general. The intermittency of a spreading process is associated with the observed fat-tailed distribution of waiting times between consecutive space steps of the CTRW. For three-wave interactions, the asymptotic spreading is non-Markovian, with Cauchy-L\'evy flights. It corresponds to the $\alpha$ value equal to $1/3$ and the $\mu$ value equal to 1. The presence of L\'evy flights signifies that the spreading is explicitly nonlocal. For four-wave interactions, nonlocality is lost in view of $\mu = 2$, while non-Markovian features attenuate to $\alpha = 1/2$. Using $\alpha$ and $\mu$, one expresses the $\nu$ value introduced in Sec.\,I as $\nu = \alpha / \mu$, where Eq.~(\ref{Pseudo}) has been considered. One gets $\nu = 1/3$ for three-wave interactions and $\nu = 1/4$ for four-wave interactions. 

\section{Discussion}

Before we proceed, we would like to make a few remarks which we believe are essential to elucidate our approach, and to further connect it to some observational and numerical evidence:

We have seen that the origin of non-Markovianity in turbulence spreading could be attributed to the presence of islands of regular dynamics having arbitrarily large sizes. The islands introduce non-ergodicity into asymptotic dynamics and favor clustering of unstable modes in phase space, with L\'evy flights representing jumps between different clusters. This picture of clustering and jumps reveals a similarity with other emergent phenomena in magnetically confined fusion plasma, among which we specifically mention the staircase self-organization \cite{DF2010,DF2015,DF2017,Horn2017,Ashour,Neg,Fang,Choi}. In a model of the plasma staircase proposed by Garbet {\it et al.} \cite{Garbet21}, wave packets of unstable modes are trapped in shear flow layers due to refraction. A staircase pattern is then idealized as a periodic radial structure of zonal shear layers that bind regions of propagating wave packets, viewed as avalanches. This connection with the plasma staircase becomes essentially clear, if one recognizes a synergetic coupling between staircasing and spreading, advocated in Ref. \cite{Nature}.    

The fact that turbulence spreading could be intrinsically nonlocal (not consistent with Fickian diffusion) finds support in flux-driven gyrokinetic simulations of L-mode tokamak plasma, reported in Refs. \cite{DF2010,DF2015,DF2017,Horn2017}. Specifically, it is found using the {\it full-f} flux-driven gyrokinetic \textsc{Gysela} \cite{Sarazin} and \textsc{XGC1} \cite{Chang} codes that the actual type of nonlocality at play is in fact the one consistent with a Cauchy-Lorentz distribution of step-sizes \cite{DF2010,Horn2017}. Lately, a Lorentzian distribution of step-sizes has been confirmed experimentally for the low-temperature plasmas of the TJ-K stellarator \cite{TJK}. Based on this evidence, we dare to say that the result $\mu \simeq 1$ is proven near marginality. This validates the nonlocal transport kernel in Eq.~(\ref{Hil}). 

Another source of evidence is the tunneling (gap spillover) problem \cite{Gurk05} according to which turbulence can overpass the locally stable regions of finite width (gaps in the growth rate profile). This topic is considered more accurately in Sec.\,VII. 

We should stress that we find nonlocality for triad interactions only, while the spreading pattern in case of four-wave interactions is local and leads to a diffusion style (Laplacian) operator in Eq.~(\ref{FDEL-L}). 
 
The reason why we do not find L\'evy flights in four-wave interactions might have a simple energy-budget explanation. Indeed, a four-wave scattering process represents a higher-order correction to the interaction Hamiltonian with respect to its three-wave counterpart, and as such has a poorer energy budget, while L\'evy flights require a strong budget \cite{Klafter,Chechkin}. In this context, it is worth noting that spreading is faster in case of three-wave interactions  (i.e., $\Delta n \propto t^{1/3}$ {\it vs}. $\Delta n \propto t^{1/4}$). 

Note, further, that we deduce the asymptotic spreading laws directly from the interaction Hamiltonian of mode-mode coupling. On the one hand, this suggests a very efficient route into a first-principle description of turbulence spreading. On the other hand, this casts doubts on the significance of dissipation processes in the turbulent domain. In fact, by placing our model on a Hamiltonian basis we have tacitly assumed that there is a time scale separation between the nonlinear spreading and dissipation times, that is, spreading occurs on much faster a time scale than dissipation destroys the turbulent motions. In hot thermonuclear plasmas, the latter condition is actually fairly well satisfied owing to strong temperature dependence of collisional viscosity \cite{Frei}.

This argument of time-scale separation is further strengthened with {\it in situ} experimental measurements of coherent structures (turbulent eddies) in a tokamak plasma. Indeed, it has been found explicitly in the edge region of the CASTOR tokamak that the lifetime of such structures is long enough to allow them to complete several full poloidal rotations before dying away \cite{Mart}. Also it has been found in the same experiments that the eddy turnover time around the density blobs is smaller than the lifetime, so that these structures could be called coherent \cite{Mart}. 

In a basic theory of fusion plasma, one describes the formation of coherent structures \cite{Huld,JJR,PLAN} using the paradigmatic Hasegawa-Wakatani model of drift-wave turbulence \cite{HW1,HW2}. Based on this, one assesses the eddy turnover time as the characteristic Rhines time pertaining to electrostatic drift waves. Likewise to its celebrated fluid analogue \cite{Rhines}, the electrostatic Rhines time \cite{Naulin-Rh} designates a characteristic time scale separating vortex motion from wave-like features and is shown to behave as inverse square root of the electrostatic drift velocity. Incorporating this insight, one infers \cite{PLA14,JPP} that strong drift-wave turbulence implies fast spreading. 

Another point of interest here concerns the origin of L\'evy flights in the spreading process. Following the analysis of Ref. \cite{PRE19}, one shows that L\'evy flights in three-wave interactions stem from the occurrence of degenerate states in otherwise a four-wave, regular interaction pattern. The degenerate states are compound states composed of two coupled waves with oppositely directed momenta and in this respect are similar to the Cooper pairs in the Bardeen-Cooper-Schrieffer picture of superconductivity in crystals \cite{Cooper}. The process is characterized by the interaction Hamiltonian 
\begin{equation}
H_{\rm int}^\prime = \frac{1}{2}\sum_{k_i, k_j} V_{k_i, k_j} \sigma^*_{k_i}\underbrace{\sigma_{k_j}^*\sigma_{k_j}}\sigma_{k_i},
\label{Coop} % Eq.~(\ref{Coop})
\end{equation}
where the brace mark singles out the degenerate state. The interaction process in Eq.~(\ref{Coop}) is in fact a three-wave process, where one of the waves involved, the compound state, behaves as a quasi-particle with zero momentum. 

These emergent considerations are made more precise in the framework of the nonlinear Schr\"odinger equation with distributed nonlinearity \cite{EPL23,PRE23}. Indeed, it is shown using a mapping procedure onto a Cayley tree that degenerate states have topological origin and correspond to self-intersections of some trees as a result of their incomplete embedding into the ambient mapping space. These self-intersections of Cayley trees produce a form of nonlocality \cite{PRE19}, which conforms to the notion of a L\'evy flight in configuration space. 

From a plasma physics perspective, the interaction Hamiltonian in Eq.~(\ref{Coop}) alludes to the zero-frequency resonance~(\ref{R0}) and in that regard describes the excitation of either zonal flows \cite{Zonal,Zonal06} or radially extended eddies or streamers \cite{Strph,Stream}. More explicitly, it is found \cite{Stream} in a simple (akin to Hasegawa-Wakatani) model of drift-wave turbulence in a tokamak that if one of the waves is a zonal flow mode (i.e., with poloidal wave number approximately zero) then radial spreading is hindered. But if one of the waves takes the form of a streamer (i.e., with poloidal wave number approximately zero) then radial spreading is enhanced. In our interpretation, the resulting spreading process corresponds to a nonlocal transport process with Cauchy-L\'evy flights and a heavy-tailed waiting-time distribution consistently with the transport model in Eq.~(\ref{Hil}), where $\alpha = 1/3$.

Furthermore, focussing on three-wave interactions, let us observe that we obtain a subdiffusive spreading {despite} the presence of L\'evy flights. Perhaps it is counter-intuitive somewhat, given that L\'evy flights are usually invoked to explain superdiffusion \cite{Klafter,Chechkin,Rest}. The point, however, is that the asymptotic spreading law is a matter of competition among the different features involved. The subdiffusive character of spreading shows that stickiness and traps take a stronger effect over the dynamics when compared to the associated nonlocal effects. As a result, the spreading process is subdiffusive at no contradiction with the fact that it is nonlocal.  

These viewpoints might be supported by experimental results of Inagaki {\it et al.} \cite{Ina} who demonstrated global hysteresis between the turbulence intensity and the local temperature gradient in the Large Helical Device L-mode plasma. Indeed, the hysteresis discloses memory in a turbulent system, suggesting that the relationship between fluxes and gradients is {\it not} instantaneous and that the behavior is intrinsically non-Markovian. That is exactly what the kinetic equations~(\ref{FDEL-L}),~(\ref{FDEL}) and~(\ref{Hil}) would imply as they incorporate memory via the Riemann-Liouville fractional derivative. Similar ideas though without the use of fractional derivatives were advocated by G\"urcan {\it et al.} \cite{PoP13}, where a simple flux-gradient relation that involves time delay and radial coupling is derived based on drift-kinetic equation.   

\section{Fractional relaxation equation}

Doing the Fourier transform of the bifractional Eq.~(\ref{FDEL}) one obtains the fractional relaxation equation \cite{Klafter,Sokolov} 
\begin{equation}
\frac{d}{d t} \hat f (k, t) =-\tau_k^{-\alpha}\,_{0}{D}_t^{1-\alpha} \hat f (k, t),
\label{Relax} % Eq.~(\ref{Relax})
\end{equation}
where $\hat f (k, t)$ denote the Fourier components of $f (n, t)$ and we have introduced 
\begin{equation}
\tau_k^{-\alpha} = |k|^{\mu}K^{\mu}_{\alpha}.
\label{Tau} % Eq.~(\ref{Tau})
\end{equation}
%$\tau_k^{-\alpha} = |k|^{\mu}K^{\mu}_{\alpha}$. 
In writing Eq.~(\ref{Relax}) we took into account that the Fourier transform of the Riesz operator~(\ref{Def+}) is $-|k|^{\mu}$, where one suppresses the imaginary unit $i^{\mu}$ following a convention used in fractional calculus (Ref. \cite{Klafter}, p. 26). 

The solution of the fractional relaxation equation~(\ref{Relax}) satisfying the initial condition $\hat f (k, t=0) = 1$ is given by the Mittag-Leffler function \cite{Klafter,Samko,Springer}
\begin{equation}
E_{\alpha} [-(t/\tau_k)^{\alpha}] = \sum_{m=0}^{\infty}\frac{[-(t/\tau_k)^{\alpha}]^m}{\Gamma(1+\alpha m)},
\label{ML} % Eq.~(\ref{ML})
\end{equation}
where $\Gamma$ denotes the Euler gamma function. For $t\gg \tau_k$, the Mittag-Leffler function $E_{\alpha} [-(t/\tau_k)^{\alpha}]$ is approximated by a power law 
\begin{equation}
E_{\alpha} [-(t/\tau_k)^{\alpha}] \simeq \frac{1}{\Gamma(1-\alpha)}(t/\tau_k)^{-\alpha},
\label{MLLT} % Eq.~(\ref{MLLT})
\end{equation}
showing that $\hat f (k, t) \simeq ((t/\tau_k)^{\alpha}\Gamma(1-\alpha))^{-1}$ for $t\rightarrow+\infty$. Assuming three-wave interactions ($\alpha = 1/3$) one has $\hat f (k, t) \propto (t/\tau_k)^{-1/3}$, while for four-wave interactions ($\alpha = 1/2$) one obtains $\hat f (k, t) \propto (t/\tau_k)^{-1/2}$. 

These theory findings are supported by results from the CASTOR tokamak \cite{Mart} described above, where it was found that the relaxation function had a power-law shape $f (\tau) \simeq (\tau/\tau_0)^{-\alpha}$, with the $\alpha$ value ranging between 0.3 and 0.5 depending on parameters of the plasma discharge and the time interval that was analyzed. We interpret this compliance with the CASTOR measurements as a confirmation that the relaxation dynamics is indeed non-Markovian and involves a long-time power-law tail consistently with the relaxation pattern in Eq.~(\ref{MLLT}). 

Another point worth noting is that the distribution of waiting times in Eq.~(\ref{WT}) could be translated into a power-law frequency distribution in accordance with
\begin{equation}
\chi_\alpha (\omega) = \chi_\alpha (\Delta t) \frac{d}{d\omega}\Delta t \propto \omega^{1+\alpha}\omega^{-2}\simeq \omega^{-(1-\alpha)}.
\label{FF} % Eq.~(\ref{FF})
\end{equation} 
More explicitly, we have $\chi_\alpha (\omega) \propto \omega^{-2/3}$ for $\alpha = 1/3$, and $\chi_\alpha (\omega) \propto \omega^{-1/2}$ for $\alpha = 1/2$. Such frequency spectra have been observed in the edge region of different tokamaks and discussed in terms of self-organized criticality \cite{Diam95,Sanchez,Sharma,Carr1,Carr2,Pol99}. Note that in both cases the frequency spectrum diverges in the infrared limit, i.e., $\lim_{\omega\rightarrow+0}\chi_\alpha (\omega) = +\infty$, emphasizing that the dynamics is non-Markovian and dominated by long correlations over time, as expected \cite{Bak1,Jensen,Works}. 

\section{Decelerating wave fronts}

In the above we have tacitly assumed that the strength of nonlinear interaction does not depend on the number of states involved. In other words, the coefficients $V_{k, -k_1, -k_2}$ were taken constant. This is actually true for small amplitudes, but it is not true in general. In magnetically confined fusion plasma, the excitation of turbulence at one radial location perturbs the state of fluctuations at the locations nearby. If those locations are found in an nonequilibrium state with population inversion (similar to active laser media \cite{Laser}), then their response to the input instability may be actually nonlinear amplifying the original perturbation. If, moreover, the process occurs near its instability threshold, then it gives birth to a chain reaction of propagating perturbation-amplification events, or avalanches \cite{Korean,Politzer,Sanchez,Sharma}. This propensity of L-mode tokamak plasma to produce avalanches has been used in Refs. \cite{PLA14,JPP} to explain the occurrence of large-amplitude transport events near the plasma edge. Also in Refs. \cite{PLA14,JPP} one finds a condition for the onset of avalanche transport in terms of the electrostatic Rhines time (which must be small compared to the instability growth time; the latter time is related with the nonadiabaticity parameter, which occurs in the Hasegawa-Wakatani model of electrostatic drift-wave turbulence \cite{HW1,HW2} and characterizes the deviation between the potential and the density fluctuations). Another class of topics and problems concerns the occurrence of self-amplifying avalanches produced by very energetic particles (e.g., intense ion beams, charged fusion products) in a tokamak \cite{Zonca05,Zonca06,Zonca15}. The phenomenon has been a subject matter of careful investigation (Refs. \cite{Heid,Zonca_RMF} for reviews; references therein). Here, we include the avalanche phenomena by permitting $V \equiv V_{k, -k_1, -k_2}$ to depend on width of the field distribution, i.e., $V = V (\Delta n)$. In a crudest setting, a proportionality relationship might be advocated, i.e., $|V| \propto \Delta n$. That means that the strength of nonlinear interaction intensifies while spreading. Combining with Eq.~(\ref{RR}) and using $A \propto |V|^2$, one gets 
\begin{equation}
(\Delta n)^2 \propto t^{1/s},
\label{Sc} % Eq.~(\ref{Sc})
\end{equation} 
where the $s$ value lies within $1/2 \leq s \leq 1$. The partial cases of Eq.~(\ref{Sc}) are a ballistic spreading $(\Delta n)^2 \propto t^{2}$ for three-wave resonances ($s=1/2$) and a diffusive spreading $(\Delta n)^2 \propto t^{1}$ for four-wave resonances ($s=1$). If $1/2 < s < 1$, then Eq.~(\ref{Sc}) predicts a superdiffusive behavior. 

From a dynamical perspective, the scaling dependence in Eq.~(\ref{Sc}) corresponds to the fractional wave equation \cite{Rest,FWE}
\begin{equation}
\frac{\partial^2}{\partial t^2} f (n, t) = K_{2-\gamma}\,_{0}{D}_t^{\gamma}\frac{\partial^{2}}{\partial n^{2}} f (n, t),
\label{FWE} % Eq.~(\ref{FWE})
\end{equation}
where $K_{2-\gamma}$ is the generalized group velocity and $_{0}{D}_t^{\gamma}$ is the Riemann-Liouville fractional derivative of the order $\gamma =  (2s - 1)/s$. We have $\gamma = 0$ for three-wave interactions and $\gamma = 1$ for four-wave interactions. The fractional wave equation~(\ref{FWE}) describes wave processes with memory in much the same way as the fractional diffusion equation~(\ref{FDEL-L}) describes subdiffusion in systems with a distribution of trapping times \cite{Rest,FWE}. In this context, the diffusive scaling $(\Delta n)^2 \propto t^{1}$ for $s=1$ has a different implication in that it characterizes the propagation of a self-decelerating wave front, not a Markov diffusion process. For $s=1/2$, no self-deceleration is to be expected, since the propagation is perfectly ballistic in this case. Mathematically, this is clear from the fact that the Riemann-Liouville derivative $_{0}{D}_t^{\gamma}$ becomes an identity operator for $\gamma \rightarrow +0$. We associate the ballistic scaling in Eq.~(\ref{Sc}) with the processes of turbulence spreading mediated by radially propagating blob-filaments of elevated ion and electron pressure \cite{Zweben,Manz,Manz20}. 

In tokamak applications, blob-filaments are magnetic-field-aligned plasma structures that are considerably denser than the surrounding background plasma and are highly localized in the directions perpendicular to the equilibrium magnetic field lines. In experiments and simulations, intermittent filaments are often formed near the boundary between open and closed field lines, and seem to arise in theory from saturation process for the dominant edge instabilities and turbulence. Blob transport is of interest from a fundamental scientific perspective, since it is a general phenomenon occurring in nearly all plasmas \cite{Ippolito}. 

In astrophysical applications, analogue blob-filaments correspond to coronal mass ejections \cite{Asch2011,Asch2016} in which a large cloud of energetic and highly magnetized plasma erupts from the solar coronal into space.   

In this study, we propose based on the scaling relation in Eq.~(\ref{Sc}) that the ballistic spreading results from triad interactions with linear amplification, i.e., $|V| \propto \Delta n$. Previous attempts to obtain a ballistic scaling for turbulence spreading in a tokamak have referred to toroidal coupling between adjacent poloidal wave numbers \cite{Garbet}. More recently, a semi-phenomenological model with bistability of turbulence intensity has been advocated \cite{Hein}, which did not use toroidal geometry to explain the ballistic spreading. Other models have employed a nonperturbative bivariate Burger's equation for transport of turbulence intensity \cite{Malkov,Malkov_NF} and a nonlinear telegraph equation for the deviation from marginality \cite{PoP13,Manz20,Kosuga}. Based on these models, a conclusion has been drawn that the heat transport in tokamaks is dominated by large avalanches with convective radial spreading \cite{Korean}. This suggestion is similar in spirit to the models of turbulence overshoot based on a complex nonlinear wave equation with feedback of wave intensity on the mode drive \cite{Zonca06,Zonca05,Zonca15,ZCh06}. In that regard, the inclusion of feedback has been shown to result in convective linear amplification of avalanches and the ballistic radial propagation of the driving source, exactly as observed in simulations \cite{Zonca05,FZ08}. If the conjecture above that $|V|$ may depend on the number of states is true, then the fractional wave equation~(\ref{FWE}) appears to be a simplest dynamical model of avalanche phenomena yet accounting for nonlinear coupling to the mode drive. In this paradigm, the use of a fractional index $\gamma =  (2s - 1)/s$ in Eq.~(\ref{FWE}) is just another way to include a time delay into the basic flux-gradient relations. That would result in a rather simple description of avalanches and explains breaking of the gyro-Bohm transport scaling in the turbulent domain, an important topic for the fusion community \cite{Korean,Nature}.      

\section{Under-barrier penetration}

We remarked in Sec.\,II that the islands of regular motion could be a ``trouble" point for chaotic systems. This is because the dynamics in the islands is isolated from the dynamics in the stochastic sea and vice versa. The presence of islands introduces non-ergodicity and$-$as a consequence$-$some non-Markovianity into the spreading process. Let us now discuss what happens at the boundary between stochastic sea and an island. 

If $T\rightarrow 0$, then the crossover from stochastic to regular dynamics must be actually very sharp (comparable to the resonance width, also about zero). An increase in $T$ implies that the stochastic motions have intensified inside the sea$-$then they would naturally result is the occurrence of a finite skin layer where turbulence intensity spills into the island. In what follows, we are interested in understanding this process in more detail. 

We begin with a notice that stochastic and regular trajectories must remain isolated also within the spillover region. This requirement is fundamental to Hamiltonian chaos \cite{Lieb} and is clear from the canonical equations~(\ref{Can}). A consequence of this is that the spillover region has an intricate fine structure, which is sandwich-like \cite{Report}: It is composed of alternating narrow stochastic layers due to the spillover of chaos and residual layers of quasi-regular trajectories. Each stochastic layer inside the island will give rise to its own skin layer, and so forth. Idealizing this picture, one shows \cite{Sagdeev} that Hamiltonian chaos has a self-similar organization and is best described by a fractal geometry. The comprehension of fractality and of multi-scale structure of phase space pertaining to chaotic motions lies at the heart of fractional dynamics (Ref. \cite{Report} for the full discussion). That said, we aim at obtaining the density of the probability to find a stochastic layer at a certain given depth inside an island. 

For this, let us model the effect of an island in terms of a potential function $\Phi (n)$, such that $\Phi (n) \equiv 0$ in the stochastic sea and $\Phi (n) > 0$ inside the island. The implication is that the island acts as a potential barrier to chaotic motions. In the above we have set the origin $n= 0$ exactly at the borderline separating chaotic trajectories inside the sea from the first regular trajectory belonging to the island. In this setting, the positive semi-axis looks towards the interior of the island, whereas the negative semi-axis looks towards the stochastic sea. It is assumed that turbulence has reached its nonlinear saturation level, hence the condition $\Phi (n) \equiv 0$ for $n \ < 0$. 

The inclusion of $\Phi (n)$ leads to the bifractional Fokker-Planck equation \cite{Klafter,Rest,Barkai}
\begin{equation}
\frac{\partial}{\partial t} f (n, t) =\,_{0}{D}_t^{1-\alpha}\Big(\frac{1}{\eta_{\alpha}}\frac{\partial}{\partial n} \Phi^\prime(n) + K^{\mu}_{\alpha} \frac{\partial^{\mu}}{\partial |n|^{\mu}}\Big) f (n, t),
\label{FFPE} % Eq.~(\ref{FFPE})
\end{equation}
where $f = f (n, t)$ is the probability density to find a stochastic trajectory at time $t$ at the position $n > 0$ inside the island, 
\begin{equation}
\int_0^{+\infty} f(n^\prime, t)dn^\prime = 1
\label{Con1} % Eq.~(\ref{Con1})
\end{equation}
due to the conservation of the total probability, $\Phi^\prime(n) \equiv d \Phi (n) / d n$, and $\eta_{\alpha}$ is the generalized friction coefficient, which carries the dimension $[\eta_{\alpha}] =$~sec$^{\alpha-2}$. If $\mu \rightarrow 2$, then $K^{\mu}_{\alpha} \rightarrow T/\eta_{\alpha}$ \cite{Rest}. Here, $T$ is thermodynamic temperature and characterizes the strength of random motions inside the layers of stochastic dynamics, and we have set the Boltzmann constant to 1 for simplicity. Note that all time-fractional kinetic equations considered in the present work, i.e., Eqs.~(\ref{FDEL-L}),~(\ref{FDEL}),~(\ref{Hil}),~(\ref{Relax}), ~(\ref{FWE}) and~(\ref{FFPE}), transform into their Markov counterparts in the limit $\alpha \rightarrow 1$. This is guaranteed by both the properties of the Riemann-Liouville fractional operator $_{0}{D}_t^{1-\alpha}$ \cite{Samko,Springer} and the reduction of the multi-scale trapping process to the regular random walk with a characteristic time step \cite{Klafter,Rest}.  
  
Equation~(\ref{FFPE}) is solved by the method of separation of variables \cite{Rest}. In fact, letting $f(n, t) = \Theta_m (t)G_m(n)$ and substituting in the Fokker-Planck equation~(\ref{FFPE}), one obtains the fractional relaxation equation for $\Theta_m (t)$, i.e.,  
\begin{equation}
\frac{d}{d t} \Theta_m (t) =-\beta_{m,\alpha}\,_{0}{D}_t^{1-\alpha} \Theta_m (t),
\label{FRE} % Eq.~(\ref{FRE})
\end{equation}
and a corresponding fractional equation for the spatial eigenfunction $G_m (n)$
\begin{equation}
\Big(\frac{1}{\eta_{\alpha}}\frac{d}{d n} \Phi^\prime(n) + K^{\mu}_{\alpha} \frac{d^{\mu}}{d |n|^{\mu}}\Big) G_m (n) = -\beta_{m,\alpha} G_m (n),
\label{Eigen} % Eq.~(\ref{Eigen})
\end{equation}
where $-\beta_{m,\alpha}$ is the eigenvalue of the Fokker-Planck operator on the left-hand side of Eq.~(\ref{Eigen}). 

Combining Eqs.~(\ref{Relax}) and~(\ref{FRE}), we identify $\beta_{m,\alpha} \simeq \tau_k^{-\alpha}$, from which $\beta_{m,\alpha} \simeq |k|^{\mu}K^{\mu}_{\alpha}$, where Eq.~(\ref{Tau}) has been considered. In the limit of extremely low frequencies (long wavelengths), we dare to set $\beta_{m,\alpha} \simeq 0$, leading to  
\begin{equation}
\Big(\frac{1}{\eta_{\alpha}}\frac{d}{d n} \Phi^\prime(n) + K^{\mu}_{\alpha} \frac{d^{\mu}}{d |n|^{\mu}}\Big) G_m (n) \simeq 0
\label{Dare} % Eq.~(\ref{Dare})
\end{equation}
for $|k|\rightarrow 0$. If $\mu = 2$ exactly, then Eq.~(\ref{Dare}) integrates to 
\begin{equation}
G_m (n) \simeq \exp [-\Phi (n) / \Phi_0],
\label{Int} % Eq.~(\ref{Int})
\end{equation}
where $\Phi_0 = \eta_{\alpha} K^{\mu}_{\alpha} = T$. One sees that turbulence may penetrate inside an island up to a characteristic depth $\ell$ for which $\Phi (\ell) \simeq \Phi_0 = T$. This is to be expected, since Eq.~(\ref{Int}) is none other than the familiar Boltzmann distribution in this case. We should stress that the notion of $T$ makes sense for $\mu = 2$ (and {\it only} for $\mu = 2$) for which L\'evy flights are absent. If $T=0$, then the penetration depth is zero. Note, in that regard, that the temperature $T$ being equal to zero corresponds to regular dynamics, which do not spread.  

If $\mu < 2$, then the notion of temperature is not well-defined (because the turbulent system with L\'evy flights is too far away from thermal equilibrium), and the spillover may be non-exponential. This is demonstrated as follows. 

One rewrites Eq.~(\ref{Dare}) with the Riesz fractional derivative in its explicit form as 
\begin{equation}
-\frac{1}{\eta_{\alpha}}\frac{d}{d n} [\Phi^\prime(n)G_m (n)] \simeq \frac{K^{\mu}_{\alpha}}{\Gamma_\mu}\frac{d^2}{d n^2} \int_{-\infty}^{+\infty}\frac{G_m (n^\prime)}{|n-n^\prime|^{\mu - 1}} dn^\prime,
\label{Exp-R} % Eq.~(\ref{Exp-R})
\end{equation}
where Eq.~(\ref{Def+}) has been used. Remembering that $\Phi (n) \equiv 0$ for $n < 0$, one is entitled to reduce the limits of improper integration on the right-hand side to the positive semi-axis only. This yields
\begin{equation}
-\frac{1}{\eta_{\alpha}}\frac{d}{d n} [\Phi^\prime(n)G_m (n)] \simeq \frac{K^{\mu}_{\alpha}}{\Gamma_\mu}\frac{d^2}{d n^2} \int_{0}^{+\infty}\frac{G_m (n^\prime)}{|n-n^\prime|^{\mu - 1}} dn^\prime.
\label{Exp-R+} % Eq.~(\ref{Exp-R+})
\end{equation}
On account of Eq.~(\ref{Con1}) one requires $\int_0^{+\infty}G_m(n^\prime)dn^\prime = 1$. Confident of this, one infers that the expression on the right-hand side of Eq.~(\ref{Exp-R+}) behaves with $n\rightarrow+\infty$ as a power law, i.e., $\sim (K^{\mu}_{\alpha}/\Gamma_\mu) n^{-(\mu + 1)}$. Applying this law, and eliminating the improper integral in Eq.~(\ref{Exp-R+}), one gets the asymptotic matching condition 
\begin{equation}
\Phi^\prime(n)G_m (n) \simeq (\eta_{\alpha}K^{\mu}_{\alpha}/\Gamma_\mu)n^{-\mu},\ \ \ n\rightarrow+\infty.
\label{Mat} % Eq.~(\ref{Mat})
\end{equation}
In view of a fundamental property of chaos to produce self-similar structures \cite{Sagdeev,Report} one might arguably propose that the potential function $\Phi (n)$ takes a power-law shape, i.e., $\Phi (n) \propto n^q$ for $n > 0$, where $q$ is a power exponent. Combining with Eq.~(\ref{Mat}), one obtains the following asymptotic behavior of the spatial eigenfunction $G_m (n)$:  
\begin{equation}
G_m (n) \propto (\eta_{\alpha}K^{\mu}_{\alpha}/q\Gamma_\mu)n^{-(\mu+q-1)},\ \ \ n\rightarrow+\infty.
\label{Mat+} % Eq.~(\ref{Mat+})
\end{equation}
It is assumed that $\Phi (n)$ is concave for all $n > 0$, with the vanishing first and second derivatives for $n\rightarrow +0$. The implication is that the exponent $q$ must be greater than 2, i.e., $q > 2$. That would ensure that there are no singularities (no fixed points) at the transition region between the sea and the island. From Eq.~(\ref{Mat+}) one sees that the decay of $G_m(n)$ is {\it algebraic}, rather than exponential. We associate the algebraic behavior in Eq.~(\ref{Mat+}) with the presence of L\'evy flights for $\mu < 2$. If $\mu\rightarrow 2$, then the normalization parameter $\Gamma_\mu$ diverges, i.e., $\Gamma_\mu = -2\cos(\pi\mu/2)\Gamma(2-\mu)\rightarrow+\infty$. This kills the power law in Eq.~(\ref{Mat+}), which by capturing locality reincarnates as an exponentially decaying $G_m (n)$ in Eq.~(\ref{Int}). If $\mu \rightarrow 1$, then $\Gamma_\mu \rightarrow 0$. In that case, one needs to redo the derivations using the Hilbert transform operator in Eq.~(\ref{Hilbert}). The end result is
\begin{equation}
G_m (n) \propto (\eta_{\alpha}K^{\mu}_{\alpha}/q\pi)n^{-q},\ \ \ n\rightarrow+\infty.
\label{Mat++} % Eq.~(\ref{Mat++})
\end{equation}
The fact that chaos can penetrate under a potential barrier with an algebraically decaying probability density function has been addressed for L-mode near-marginal tokamak plasma in Ref. \cite{PRE18}. The phenomenon was subsequently demonstrated in gyrokinetic simulations using the \textsc{Gysela} code \cite{PRE21}. It has been discussed \cite{PRE18,PRE21} that an algebraic form in Eq.~(\ref{Mat+}) gives rise to a weak localization of Cauchy-L\'evy flights, where by {\it weak} localization one means a situation according to which a L\'evy flight acquires finite second moments in the presence of a potential field \cite{Chechkin,Rest,Ch2004,Bimod}. In that context, one shows, following Refs. \cite{Chechkin,PRE18,Bimod}, that there exists a minimal $q$ value, which weakly localizes a Cauchy-L\'evy flight. The demonstration consists in requiring that the integral $\int G_m (n^\prime)n^{\prime2}dn^\prime$ converges at infinity. This yields   
\begin{equation}
q > 4-\mu,
\label{Conv} % Eq.~(\ref{Conv})
\end{equation}
from which $q_{\min} = 3$, where the exponent $\mu = 1$ has been applied. (If one is a purist and wants to be careful with limits, one adheres to $q_{\min} = \lim_{\epsilon\rightarrow +0} 3 + \epsilon$.) Note that the condition $q > 2$ has been satisfied. 

Another issue worth noting is that feedback dynamics near marginality stabilizes a Cauchy-L\'evy flight {exactly} at its delocalization point. The explanation \cite{PRE18,PRE21} refers to self-organized criticality \cite{Bak1,Bak2,Jensen}, through which complex systems find their critical states. Given this insight, one applies $q=q_{\min} = 3$ in Eq.~(\ref{Mat++}) to get
\begin{equation}
G_m (n) \propto (\eta_{\alpha}K^{\mu}_{\alpha}/q\pi)n^{-3},\ \ \ n\rightarrow+\infty.
\label{Mat3} % Eq.~(\ref{Mat3})
\end{equation}
This is the end result. Naturally we associate the power-law drop-off in Eq.~(\ref{Mat3}) with three-wave interactions near a marginally stable state, consistently with the analysis of Sec.\,II. 
 
We should stress that the results~(\ref{Mat+}),~(\ref{Mat++}) and~(\ref{Mat3}) are at odds with a suggestion of G\"urcan {\it et al.} \cite{Gurk05} that the spillover of turbulence into a stable region goes on an exponentially decaying probability density. The discrepancy occurs because G\"urcan {\it et al.} used the diffusion Ansatz in postulating their diffusion-reaction equation, while in our study we deliberately avoided invoking this Ansatz. Yet, we recover an exponential decay in the parameter range of four-wave interaction, for which $\mu = 2$.   

We conclude with a warning that the algebraic forms in Eqs.~(\ref{Mat+}),~(\ref{Mat++}) and~(\ref{Mat3}) indicate an enhanced probability of turbulence penetration under a transport barrier$-$potentially posing a threat to safety of the tokamak plasma operations \cite{PRE18}. The main risk here is owed to the occurrence of large-amplitude and extreme events \cite{PRE18,PRE21,Asch2016,Sornette,Sornette_2012,Sornette_PRL,Eliazar,Sharma,CSF21}, whose origin is fully dynamical and related with the emergent behavior of systems with many interacting degrees of freedom \cite{Asch2011,Jensen,Works,Jensen2}. The comprehension of such events in magnetic confinement fusion is in its infancy.   

\section{Summary and conclusions}

We have proposed a theory of turbulence spreading based on fractional kinetics. Fractional versions of the diffusion, relaxation, wave, and Fokker-Planck equations have been considered. 

The use of fractional-derivative equations is motivated with non-ergodic character of the asymptotic spreading. The non-ergodicity is associated with the presence of islands of regular motion having arbitrarily large sizes. The role of the islands is that they hamper prompt mixing of phase-space trajectories and by doing so introduce long-time memory into the spreading process, leading to anomalously slow (powerlaw-like) decay of the time correlations. As a result, the asymptotic spreading is weakly chaotic, with vanishing Lyapunov exponents$-$and not strongly chaotic, as most theories based on the conventional Fokker-Planck schemes would imply.  

The spreading occurs because the unstable modes couple together via the resonant wave-wave interactions giving rise to layers of stochastic dynamics. The kinematics of the spreading process depends on whether the resonances occur between three or four waves: 

In a three-wave picture of interactions, the spreading process is shown to be non-Markovian, with Cauchy-L\'evy flights. The effect of L\'evy flights is that they introduce explicit nonlocality into the dynamics. It is understood that the familiar Fick's law that fluxes at a point are produced by gradients at the {\it same} point does not apply in this case.

In a four-wave picture instead, the asymptotic spreading appears to be local in the sense it does not involve L\'evy flights. Yet, it is non-Markovian as a result of non-ergodicity and weak mixing in the limit $t\rightarrow+\infty$.  

In either three- or four-wave interaction scenario, the asymptotic spreading is found to be subdiffusive (due to the overall non-Markovianity of the dynamics). It is somewhat faster in case of three-wave interactions (owing to the presence of L\'evy flights). 

We have collected these and other relevant findings in Table~1, from which the similarities and differences between three- and four-wave coupling processes are clear.

\begin{table}[t]
\begin{center}
\begin{tabular}{p{3.0cm}p{2.5cm}p{2.5cm}} \hline \hline
Exponent & Three-wave & Four-wave \\ 
\hline \hline
$s$,\ $1/2 \leq s \leq 1$ & 1/2 & 1 \\
$\alpha = s/(s+1)$ & $1/3$ & $1/2$ \\
$\mu = 2s$ & $1$ & $2$ \\ 
$\nu = \alpha / \mu$ & $1/3$ & $1/4$ \\
$\gamma = (2s - 1)/s$ & $0$ & $1$ \\
$q$,\ $q \geq 4-\mu$ & $3$ & Not applicable\footnote{This case corresponds to strong localization and is characterized by the exponential drop-off in Eq.~(\ref{Int}).} \\
\hline \hline
Equation/property & Three-wave & Four-wave \\ 
\hline \hline
$\langle(\Delta n)^2\rangle$ & $\propto t^{2/3}$ & $\propto t^{1/2}$ \cite{PRE17,PRE23} \\
$\chi_\alpha (\Delta t)$ & $\propto (\Delta t)^{-4/3}$ & $\propto (\Delta t)^{-3/2}$ \\
$\chi_\mu (|\Delta n|)$ & $\propto |\Delta n|^{-2}$ & Gaussian \\
$\chi_\alpha (\omega)$ & $\propto \omega^{-2/3}$ & $\propto \omega^{-1/2}$ \\
$\hat f (k, t)$ & $\propto (t/\tau_k)^{-1/3}$ & $\propto (t/\tau_k)^{-1/2}$ \\
$G_m (\Delta n)$ & $\propto (\Delta n)^{-3}$ & $\exp [-\Phi (\Delta n) / \Phi_0]$ \\
\hline \hline
Transport equation & Eq.~(\ref{Hil}), $\alpha = \frac{1}{3}$ & Eq.~(\ref{FDEL-L}), $\alpha = \frac{1}{2}$ \\
Nonergodic & Yes & Yes \\
Non-Markovian & Yes, $\alpha = \frac{1}{3}$ & Yes, $\alpha = \frac{1}{2}$ \\
Trappings & Yes & Yes \\
Nonlocal &Yes & No \\
L\'evy flights & Yes, $\mu = 1$\footnote{Special value corresponding to Cauchy-L\'evy flights \cite{Ch2004}.} & No, $\mu = 2$\footnote{Gaussian limit of the Riesz fractional derivative \cite{Rest,Chechkin}.} \\
Fick's law & No\footnote{On account of both non-Markovianity and nonlocality.} & No\footnote{On account of non-Markovianity only, otherwise local relationships may apply.} \\
Wave fronts & Ballistic\footnote{In case of the direct proportionality $|V| \propto \Delta n$.} & Diffusive\footnote{In case of the direct proportionality $|V| \propto \Delta n$.} \\
\hline \hline
\end{tabular}
\end{center}
\caption{A summary of results and comparison between three- and four-wave interaction patterns. One sees that the asymptotic spreading is faster in case of triad interactions for which it also involves L\'evy flights (on an equal footing with non-Markovian features).} \label{tab1}
\label{default}
\end{table}

A close inspection of Table~1 suggests a set of unique signatures or fingerprints of three-wave interactions: a relatively fast asymptotic spreading complying with an $\Delta n\propto t^{1/3}$ scaling; an explicitly nonlocal behavior with Cauchy-L\'evy flights; and an algebraic, rather than exponential, tunneling pattern. The fingerprints of four-wave interactions are, on the contrary, a slower spreading conforming to an $\Delta n\propto t^{1/4}$ behavior; the absence of flights; and an exponentially decaying density of the probability to spill over a barrier. Both spreading patterns appear to be non-Markovian, with a distribution of trapping times.  

That said, one wonders based on which principles the actual system with linear instability chooses a three- or four-wave interaction pattern or both. {\it A priori} this question is not easy to answer. From an energy-budget viewpoint, three-wave interactions might be preferred as they correspond to a lower-order correction to $H_0$, whereas four-wave interactions correspond to a higher-order correction. Not surprisingly, it is found using the Hasegawa-Wakatani model of drift-wave turbulence that triad interactions are the most effective in turbulence spreading \cite{Stream}. 

A more sophisticated answer refers to the dispersion relation $\omega_i = \omega_i (k_i)$, i.e., to the exact instability at play. In plasmas, the concrete form of the $\omega_i = \omega_i (k_i)$ dependence may be actually quite complex and nonlinear. It may produce a decaying spectrum with regard to e.g., three-wave interactions and not four-wave interactions; or {vice versa}; or both; or neither of these. It may result in a decaying spectrum in 2D and not in 1D, etc. If the function $\omega_i = \omega_i (k_i)$ is known, then one may predict the type of interaction by solving the respective systems of equations~(\ref{R3})~and/or~(\ref{R4a})$-$(\ref{R4b}) using graphical methods. This approach is elucidated in Ref. \cite{Sagdeev}. 

If both three- and four-wave interactions are allowed by the dispersion relation, then the actual spreading rate is determined by the three-wave interactions (because the spreading process is much faster in that case).   

In this study, we committed ourselves to clear distinction between three- and four-wave interaction processes. Yet, one might be interested in obtaining a unifying picture, which interpolates between the two regimes. That turns out to be possible, if one relies on a theoretical scheme of the nonlinear Schr\"odinger equation with subquadratic power nonlinearity \cite{PRE19,PRE23,EPL23,PRE21}. In that description, three-wave interactions appear to be a singular case as they correspond to the nonanalytical modulus function and not to the familiar modulus squared, as with four-wave interactions. It is shown, accordingly, that this non-analiticity reflects the presence of degenerate states in an otherwise regular four-wave interaction pattern and that such states give rise to L\'evy flights. The demonstration is straightforward, but lengthy. It uses a mapping procedure onto a sequence of Cayley graphs with odd coordination numbers. For each coordination number, one solves a system of Diophantine equations through which the selection rules for L\'evy flights can be inferred. The end result is that unlimited spreading may occur if the interaction process involves {\it at least} three waves (or more), and is forbidden otherwise.            

We have seen in the above that the asymptotic spreading is characterized by a bifractional diffusion equation with competition between fat-tailed trapping-time and step-size (for three-wave interactions only) distributions. In this paradigm, trappings result from clustering of unstable modes in phase space \cite{Iomin,PRE17} and mathematically correspond to the action of a binding potential of the Lennard-Jones type. In a similar spirit, L\'evy flights represent the jumps {between} different clusters. The inclusion of L\'evy flights implies that turbulence can overpass the domains of regular dynamics and by doing so emerge in locations that are disconnected from the original location$-$exactly as observed in simulations \cite{Gurk05}. 

Using the idea of Lennard-Jones potential, we obtained the exponents of the trapping-time and step-size distributions without turning to numerical simulations. In that regard, our theory predicts the exponents of the fractional transport equation {\it self-consistently} from the model itself. That said, the results presented in this work prove against simulations, with comprehensive numerical evidence reported in e.g., Refs. \cite{DF2017,Flach,Skokos,Ivan,Senyange,PRE21}, just to mention some.  

The effects of stickiness, weak chaos and trapping of dynamical trajectories are further singled out by taking Fourier transform of the asymptotic transport equation, leading to the fractional relaxation equation and the ubiquitous Mittag-Leffler relaxation pattern \cite{Sokolov,Springer}. We have seen that the Mittag-Leffler function correctly reproduces the observed power-law decay \cite{Mart} of the autocorrelation function in the edge region of the CASTOR tokamak. Another milestone is the finding of Ref. \cite{Ina} that there exists a global hysteresis between the turbulence intensity and the local temperature gradient, which strongly suggests the application of kinetic equations with memory.      

We should stress that our approach is based on the interaction Hamiltonian of the resonant mode-mode coupling and in this sense does not involve phenomenological or heuristic assumptions. In fact, by starting from a basic Hamiltonian of mode-mode interaction we attempted a theory of turbulence spreading without specifying the kind of instability behind (e.g., electrostatic drift wave, ion temperature gradient, interchange, etc.). In this fashion, we have focussed our analysis on the generic mechanism of turbulence spreading regardless of the very specific instability model and of the specific turbulence type. This is why our study is distinct from previous works. 

The main message of our theory is that the asymptotic spreading has complex microscopic organization, is nondiffusive and intermittent in general. In fusion literature, there exists an outstanding evidence, both experimental and numerical, that the radial (cross-field) transport may be intermittent, with avalanches and bursts (e.g., Refs. \cite{Ben,Politzer,Tokunaga,Zweben,Manz,Manz20,Ippolito}; Refs. \cite{Rev15,Sanchez} for reviews). We have seen that intermittency originates from non-ergodicity of turbulence spreading and is something related with inhomogeneous, not space-filling turbulence. In fact, non-ergodicity favors the occurrence of large-long dynamical fluctuations \cite{Report,Edel}$-$avalanches and bursts$-$whose probabilities might not be exponentially small. This is clear from the power-law step-size distribution in Eq.~(\ref{LW}) leading to L\'evy flights. The occurrence-frequency (waiting-time) distribution of bursts is given by Eq.~(\ref{WT}) and is motivated with stickiness of dynamical trajectories in phase space for $t\rightarrow+\infty$. 

The fact that the asymptotic transport is found to be nondiffusive, with a waiting-time and bursts statistics, breathes new life into the old work of Townsend \cite{Town} who found from a study of the turbulent wake behind a cylinder in hydrodynamics that ``the use of a diffusion coefficient to describe the transport of turbulent energy is not justified, and that energy diffusion is a process independent of momentum diffusion" (Ref. \cite{Town}, p. 133). Also Townsend remarks that to ``remove this difficulty, it is not sufficient to consider the effects of intermittency." From a nowadays perspective it would appear that the results of Townsend have a strong taste for fractional dynamics \cite{Report,Klafter,Sokolov,Rest,Nat3} and in that sense provide indirect support for the theory approaches employed in the present work. 

Developing these viewpoints, we have encountered a situation according to which the strength of nonlinear interaction may depend on width of the field distribution (i.e., nonlinearity intensifies while spreading). We have seen that a nonlinearity of the kind explains the ballistic spreading and the birth of avalanches naturally without involving toroidicity effects. The implication is that in our model avalanches are permitted already in a cylinder, if a certain generic condition \cite{PLA14} using the Rhines time is satisfied. In this context, our model differs from the model of Garbet {\it et al.} \cite{Garbet} which required a sort of toroidal coupling to explain the ballistic spreading.  

A generalization of the ballistic spreading corresponds to a superdiffusive, sub-ballistic spreading, which could be understood as a ballistic spreading with memory. We characterized this class of spreading phenomena using the fractional wave equation \cite{Rest,FWE}. The latter equation interpolates between diffusive and ballistic scalings and is similar in spirit to the fractional diffusion equation describing subdiffusion \cite{Klafter,Sokolov}. It is worthy to emphasize that the observation of a diffusive scaling does not imply by itself that the spreading process is diffusive. It might well be a sub-ballistic spreading with time delay obeying the fractional wave equation, with $\gamma = 1$ \cite{Rest,FWE}.        

Finally, we addressed a tunneling problem for turbulence spreading and saw that it leads to an exponential (Anderson like) localization in case of four-wave resonances and to an algebraic (weak) localization in case of three-wave resonances. In the latter (three-wave interaction) case, the probability of under-barrier leakage of turbulence is greatly enhanced. In fact, the fractional Fokker-Planck equation~(\ref{FFPE}) suggests that the density of the probability decays under a barrier as inverse cube of distance. This is in agreement with the ``comb" model of Ref. \cite{PRE18}, while is in contrast with the diffusion-reaction approach of Refs. \cite{Gurk05,Gurk06}. A lesson to be learned is that dynamical chaos might escape localization much easier than it was thought before. 

The idea of weak localization finds support in the results of gyrokinetic modeling of L-mode plasma reported in Refs. \cite{DF2010,Horn2017,PRE21}. Indeed, it has been found in those simulations that turbulence spreading might both result from$-$and via complexity couplings close to marginality also result in$-$the rise of transport barriers, and that the distribution of step-sizes of under-barrier propagation is consistent with the theoretical concept of Cauchy-L\'evy flight \cite{PRE18,PRE21}. 

From a somewhat more general perspective, the comprehension of turbulence spreading and its cross-interaction with transport barriers  suggests a paradigm shift \cite{Nature} according to which the efficient plasma performance in a tokamak is a compromise among turbulence, turbulent transport and transport barriers$-$controlled nonlocally by fluxes of turbulence activity near a globally critical state which is self-organized. Research over these topics poses a challenging task for future work.

\acknowledgments
Thanks to Guilhem Dif-Pradalier and Alexander Iomin for the pleasures of working with them, and to Peter Leth Christiansen, Patrick Diamond, Carlos Hidalgo and Xavier Leoncini for discussions. The authors acknowledge fruitful interactions with the participants in the Festival de Th\'eorie 2022, Aix-en-Provence, France. A.V.M. is grateful to Civilingeni{\o}r Frederik Christiansens Almennyttige Fond, for financial support, enabling to initiate this work at DTU in December 2021. Also A.V.M. acknowledges hospitality and financial support at the International Space Science Institute (ISSI, Bern, Switzerland), where this paper was developed. Part of the work has been carried out within the framework of the EUROfusion Consortium and has received funding from the Euratom research and training programme 2014-2018 and 2019-2020 under grant agreement No 633053. The views and opinions expressed herein do not necessarily reflect those of the European Commission. 
%Part of the work has been carried out within the framework of the EUROfusion Consortium and funded by the European Union via the Euratom Research and Training Programme (Grant Agreement No 101052200$-$EUROfusion). Views and opinions expressed are however those of the author(s) only and do not necessarily reflect those of the European Union or the European Commission. Neither the European Union nor the European Commission can be held responsible for them. 

%
%
%
%


\end{document}